\documentclass[aps,prl,twocolumn,showpacs]{revtex4} \usepackage{amsmath}
\usepackage{graphicx} \usepackage{color} \definecolor{gray}{RGB}{191,191,191}

\begin{document}

\title{Absence of metallicity in K-doped picene: Importance of electronic
correlations}

\author{Andreas Ruff}
\affiliation{Physikalisches Institut, Universit\"at W\"urzburg, Am Hubland, 97074 W\"urzburg, Germany}

\author{Michael Sing}
\affiliation{Physikalisches Institut, Universit\"at W\"urzburg, Am Hubland, 97074 W\"urzburg, Germany} 

\author{Ralph Claessen}
\affiliation{Physikalisches Institut, Universit\"at W\"urzburg, Am Hubland, 97074 W\"urzburg, Germany}

\author{Hunpyo Lee}
\affiliation{Institut f\"ur Theoretische Physik, Goethe-Universit\"at Frankfurt, 60438 Frankfurt am Main, Germany} 

\author{Milan Tomi{\'c}}
\affiliation{Institut f\"ur Theoretische Physik, Goethe-Universit\"at Frankfurt, 60438 Frankfurt am Main, Germany} 

\author{Harald O. Jeschke}
\affiliation{Institut f\"ur Theoretische Physik, Goethe-Universit\"at Frankfurt, 60438 Frankfurt am Main, Germany} 

\author{Roser Valent\'{i}}
\affiliation{Institut f\"ur Theoretische Physik, Goethe-Universit\"at Frankfurt, 60438 Frankfurt am Main, Germany}

\date{\today}

\begin{abstract} Potassium-doped picene (K$_x$picene) has recently
  been reported to be a superconductor at $x=3$ with critical
  temperatures up to 18 K. Here we study the electronic structure of
  K-doped picene films by photoelectron spectroscopy and {\it ab
    initio} density functional theory combined with dynamical
  mean-field theory (DFT+DMFT). Experimentally we observe that, except
  for spurious spectral weight due to the lack of a homogeneous
  chemical potential at low K-concentrations ($x \approx 1$), the
  spectra always display a finite energy gap. This result is supported
  by our DFT+DMFT calculations which provide clear evidence that
  K$_x$picene is a Mott insulator for integer doping concentrations
  $x=1$, 2, and 3.  We discuss various scenarios to understand the
  discrepancies with previous reports of superconductivity and
  metallic behavior. 
\end{abstract}

\pacs{71.20.Tx,79.60.Fr,71.15.Mb,71.27.+a}


\maketitle

\begin{figure}[t] \includegraphics[width=0.45\textwidth]{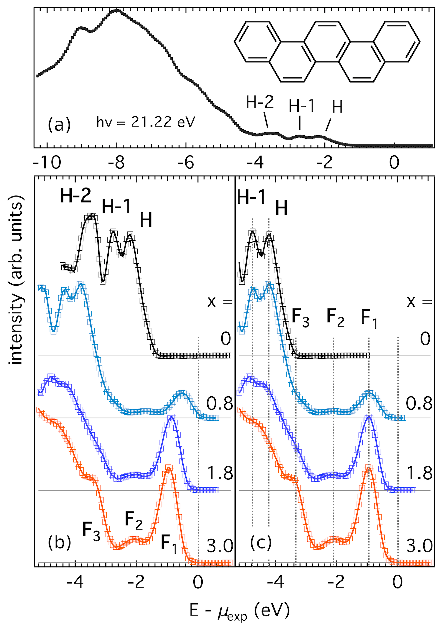}
  \caption{(a) Valence band spectrum of a pristine picene film.  H,
    H-1, and H-2 label the highest occupied molecular orbitals of the
    pristine film.  (b) Low-energy part of the valence band spectrum
    showing the evolution with K-doping for various near-integer
    concentrations. The doping-induced features in the former bandgap
    are assigned by F$_1$, F$_2$, and F$_3$. All energies as measured
    relative to the experimental Fermi energy $\mu_\text{exp}$. (c) Same spectra, 
    but energetically aligned at the spectral features of the
    $x=3$ spectrum, for which the existence of an equilibrium chemical
    potential can safely be assumed (see text for
    details).}\label{UPS} \end{figure}

Since the discovery of superconductivity in layered organic charge transfer
salts in the late 1970s \cite{Toyota2007}, a large variety of organic
superconductors have been reported with critical temperatures ranging from a few
Kelvins to values as high as 38~K in the case of alkali metal-doped fullerenes
\cite{Palstra1995}. Whether superconductivity is driven by electron-phonon
interactions or by electronic correlations has been a subject of intense debate
in the last decade, and indeed a significant fraction of these systems are
presently being discussed as correlated materials~\cite{Capone2002,Powell2011}.
Recently, a pronounced Meissner-Ochsenfeld-effect indicating a transition temperature
of 18~K was observed in polycrystalline powder samples of potassium-doped
picene~\cite{Mitsuhashi2010}. Moreover, the recent observation of zero resistivity below 7~K 
supports these findings~\cite{Teranishi2013}. The parent compound picene is a wide band-gap
semiconductor (gap $\sim$ 3.3~eV~\cite{Okamoto2008}) consisting of planar
molecules with five aromatic benzene rings (see inset in Fig.~\ref{UPS}). The
observation of superconductivity upon K intercalation has also been reported in
other aromatic semiconductors, viz., phenanthrene, coronene and
dibenzopentacene \cite{Wang2011,Kubozono2011,Xue2012}.

In a simple one-electron picture, K-intercalation is expected to induce electron
transfer into the lowest unoccupied molecular orbital (LUMO) of picene, leading
to a partially filled, i.e., metallic conduction band which at low
temperatures would host the superconductivity, as suggested in various
theoretical studies
\cite{Kosugi2009,Roth2010,Andres2011,Kosugi2011,Cudazzo2011,Kubozono2011,
Subedi2011,Casula2011,Kato2011,Casula2012}.
However, such calculations include electronic correlations only at the level of the
exchange-correlation functional considered within density functional theory
(DFT) and neglect electron-phonon interactions. Both effects can be quite
pronounced in molecular solids \cite{Claessen2002}. 
In fact, the importance of electron-electron correlations in K$_3$picene
was pointed out by Giovannetti {\it et al.}~\cite{Giovanetti2011} and Kim {\it et al.}~\cite{Kim2011} 
who found an insulating state with antiferromagnetic order
 from {\it ab initio} calculations. Experimentally, the picture
of the electronic structure has remained controversial: while one photoemission
study reported the appearance of metallic spectral weight at the Fermi level for
$x=1$ \cite{Okazaki2010}, other authors \cite{Mahns2012,Caputo2012} failed to observe a
metallic Fermi edge for a wide concentration range from $x=0.22$ to $3.6$ and
speculated about the possible role of correlations and polaronic
effects~\cite{Mahns2012}.

In view of this unsettled situation we have investigated the
electronic structure of K$_x$picene in a joint experimental and
theoretical effort. For this purpose, extensive photoemission
measurements on pristine and potassium-doped picene films have been
perfomed, accompanied by {\it ab initio} density functional theory
(DFT) calculations in combination with dynamical mean field theory
(DFT+DMFT). Contrary to the results of Ref.~\onlinecite{Okazaki2010},
both experimental and theoretical results confirm the absence of
metallic behavior for any composition of K$_x$picene with integer
$x=1,2,3$. This insulating behavior is attributed to strong on-site
(i.e., intramolecular) correlations as inferred from an
analysis of the DFT+DMFT calculations.

Picene films have been grown in ultrahigh vacuum by thermal
evaporation out of a Knudsen cell on clean natively oxidized Si
substrates. Subsequent doping was achieved by evaporation of potassium
from commercial SAES\textsuperscript{\textregistered} dispensers onto
the picene films at room temperature. \textit{In situ} photoemission
spectra were recorded at room-temperature with a total energy
resolution of 0.1 and 0.7~eV using He~I (21.2~eV) and monochromatized
Al~K$_{\alpha}$ (1486.6~eV) radiation for valence band and core level
spectroscopy, respectively. All experiments have been performed at a base
pressure of $1\cdot10^{-10}$~mbar, which during picene and K deposition did 
not rise above $5\cdot10^{-9}$~mbar and $8\cdot10^{-10}$~mbar, respectively. 
The relative intensity of the C$\,1s$ and K$\,2p$ core levels was used to determine 
the doping concentration (see also Supplemental Material). All valence band spectra have been
corrected for He I$_{\beta}$ satellite emission.

Bulk picene crystallizes in a monoclinic unit cell with the molecules
arranged in a herringbone structure \cite{De1985}. As typical for
planar molecules on inert substrates, our picene films grow in a 3D
island mode (Vollmer-Weber type growth), with the long axis standing
out from the surface in order to minimize film-substrate
interaction. The c-axis parameter determined from x-ray diffraction is
13.51~\AA, very close to its value in bulk material
\cite{De1985,Mitsuhashi2010} (see Suppl. Mat. for further
information).  Due to strong K-oxidation under ambient atmosphere we
have not been able to measure reliable structural parameters for the
K-doped films.

On the theoretical side the crystal structure of K$_x$picene ($x=1,2,3$)
has been determined by structural relaxation within DFT using the
generalized gradient approximation (GGA). The projector augmented wave
basis was used as implemented in
VASP~\cite{Kresse1993,Kresse1996,Bloechl1994,Kresse1999}, with
additional van der Waals corrections included in the relaxation
process. The subsequent analysis of the electronic structure was
performed with the full potential local orbital code
FPLO~\cite{picenestructure}.

Figure~\ref{UPS}(a) shows the valence band spectrum of a pristine picene film. It is fully consistent with previous photoemission studies \cite{Xin2012,Okazaki2010,Roth2010}, with the highest occupied molecular orbital (HOMO) and following bands clearly resolved at  -2.2 (H), -2.8 (H-1) and -3.5~eV (H-2) relative to the experimental Fermi energy ($\mu_{\text{exp}}$, determined by the Fermi cut-off of a sputtered gold foil). 
Figure~\ref{UPS}(b) displays the evolution of the spectra with K-doping from $x=0$ to 3 as measured. In agreement with previous work the doping results in additional features (F$_1$, F$_2$, F$_3$) in the former bandgap and an apparent shift of the overall valence band to higher binding energies. This shift is also seen in the C$\,1s$ core level spectrum (not shown here). 

We first discuss the origin of the additional spectral features. In
analogy to other alkali metal-doped organic semiconductors \cite{Ramsey1994,
  Koch2003} F$_3$ is assigned to the relaxed HOMO, induced by the
structural distortion experienced by the doped picene molecules. Its
intensity clearly increases with doping, while the strongly decreasing
(but still visible) contribution of the H and H-1 peaks in
Fig.~\ref{UPS} reflects the diminishing fraction of uncharged
molecules upon doping. The weak feature F$_2$ can be shown to result
from the oxidation of surface K atoms with oxygen present in the
residual gas of the ultrahigh vacuum system
\cite{Merkel1993,Ruff2012}. Therefore, it is a purely extrinsic
structure not relevant for the following discussions. Finally, the
most prominent in-gap feature F$_1$ is attributed to a
(non-rigid-band-like) filling of the former lowest unoccupied
molecular orbitals (LUMOs), as suggested by the theoretical calculations
discussed below.

Next we turn to the evolution of the apparent binding energies in the
valence band spectra. Figure~\ref{UPS}~(b) shows that for $x=0.8$ the new
peak F$_1$ sits very close or even at the experimental Fermi level,
reminiscent of the spectrum reported by Okazaki \textit{et al.}
\cite{Okazaki2010} for $x=1$, while for larger doping concentration it
clearly moves away from $\mu_\text{exp}$, leaving a (half)gap of $\approx
0.5$ eV. However, the near-Fermi level position of F$_1$ around $x=1$
should not be mistaken for a metallic Fermi edge. First of all, its
spectral onset is far too broad to account for a sharp Fermi edge,
even considering thermal and instrumental broadening. Moreover, for
smaller concentrations between $x=0$ and 1 the spectral weight of peak
F$_1$ is even observed \textit{well above} the experimental Fermi
energy (see Fig.~S2 in the Suppl. Mat.). Such curious behavior was
observed also in other alkali metal-doped organic semiconductor films and
was established to result from a lack of thermodynamic equilibrium
originating from the weak intermolecular van der Waals coupling
\cite{Koch2003}. In such a situation, charge fluctuations are strongly
suppressed, thereby preventing the formation of a well-defined
chemical potential. The ionic charge transfer induced by alkali metal-doping
restores the equilibrium. Because in our case the doping is achieved
by K-evaporation \textit{after} film growth, the doping concentration
will decrease from surface to substrate (note that the $x$-values
determined from core level spectroscopy are only valid for a thin
surface layer given by the photoemission probing depth of a few
nm). Only for sufficiently strong overall doping full equilibrium
within the organic film and between film and substrate is achieved.

In our case we observe saturation of the doping-induced energy shifts
at $x \approx 3$, which we identify with equilibrium, i.e.,
the energies of the $x=3$ spectrum can be safely referred to the
experimental Fermi energy. Consequently, we display in
Fig.~\ref{UPS}(c) all spectra aligned at the position of the unrelaxed
H and H-1 peaks of the $x=3$ film, which incidentally fixes also the
doping-induced features at constant energies. This representation nicely
shows the evolution of in-gap spectral weight upon doping, with none
of the spectra displaying any metallic Fermi edge. The observation
that even the $x=3$ phase, for which superconductivity has been
claimed, remains gapped is most surprising, but in agreement with two
previous reports~\cite{Mahns2012,Caputo2012}. As one possible
explanation polaronic effects have been suggested~\cite{Mahns2012},
which may cause a pseudogap at the Fermi level as has been observed,
e.g., in oxides \cite{Schrupp2005}. However, such a pseudogap
typically exhibits an exponential decay towards the Fermi energy
\cite{Alexandrov1992}, which is not consistent with the present
observation. Alternatively, the gap may originate from strong
electronic correlations. In fact, in a Hubbard model-type description
the intramolecular Coulomb energy $U$ in organic solids is often
comparable or even larger than the bandwidth from intermolecular
hopping \cite{Claessen2002,Nomura2012}, which can drive partially
filled bands into a Mott insulating state.

\begin{figure}[t] \includegraphics[width=0.45\textwidth]{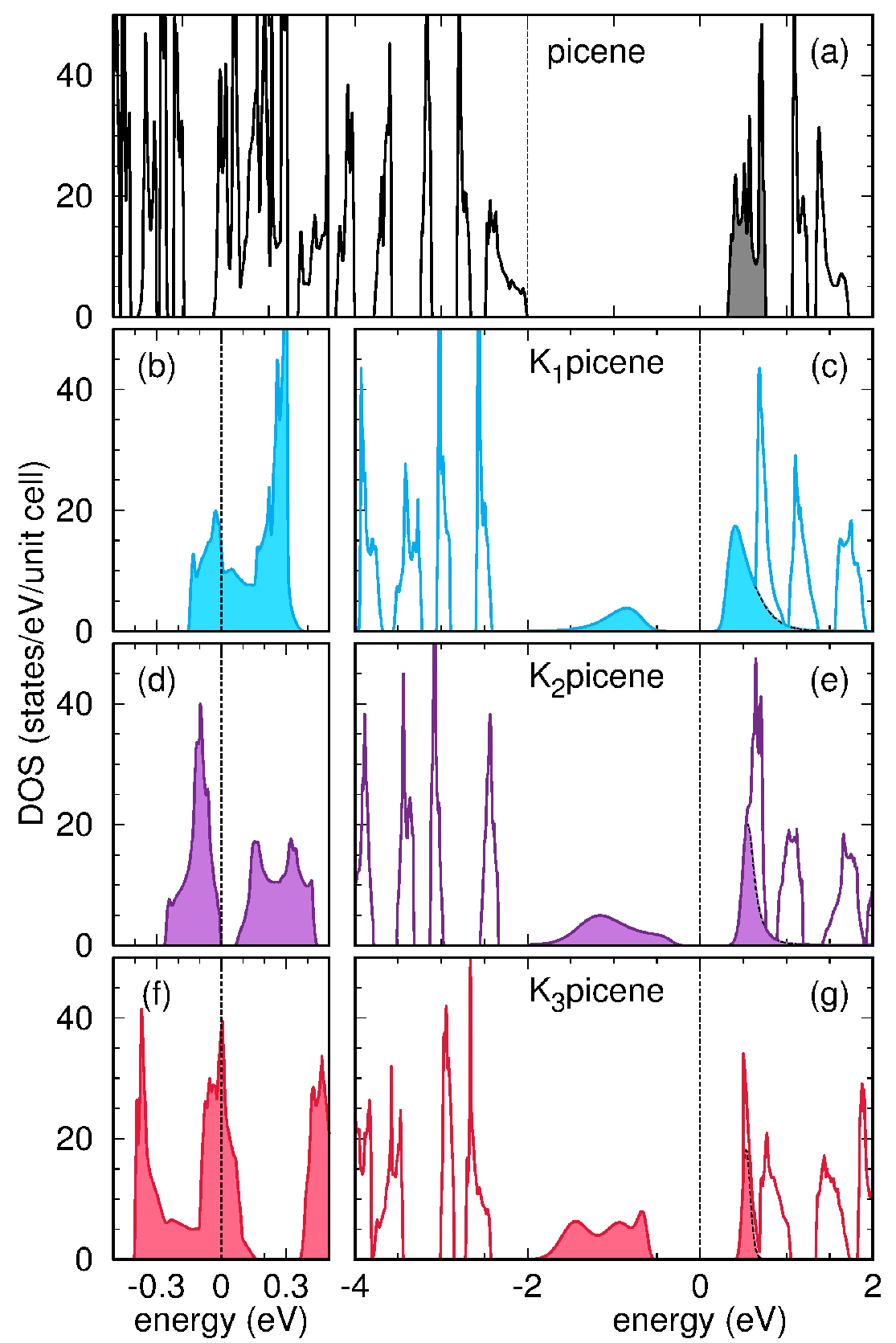}
  \caption{Density of states of pristine and K doped picene. (a) DFT
    DOS for pristine picene with the part of the spectrum arising from
    LUMO and LUMO+1 levels of picene marked by gray-shading. (b), (d),
    (f) DFT DOS of the region close to the Fermi level for K$_1$picene,
    K$_2$picene and K$_3$picene, respectively. (c), (e) and (g)
    Combined DFT+DMFT DOS of the same compounds. For details see
    text.}\label{theory} \end{figure}

In order to further explore the latter scenario, we have studied the
electronic properties of the DFT-relaxed crystal structures of
K$_x$picene ($x=1,2,3$) by a combination of DFT+DMFT calculations,
which include correlation effects beyond the GGA approximation in
DFT. We start first with the DOS for pristine and K$_x$picene obtained
within GGA (Fig.~\ref{theory}~(a), (b), (d) and (f)). We observe a
metallic state for odd $x$ as well as a band gap for $x=2$ as has also
been found in previous DFT calculations.

In DFT+DMFT the kinetic part $H_K$ of the Hamiltonian
(Eq.~\ref{eq:hami}) is obtained from the DFT calculations by
considering the tight-binding approach.  Since the DFT results
indicate some hybridization between the bands derived from LUMO and
LUMO+1 of the pristine picene molecules, we consider for the DMFT
calculations a two-orbital Hubbard model with a Coulomb interaction
$U$ and Hund's coupling $J_z$ as well as on-site hopping parameters
$t_{\perp}$ between molecular orbitals on the same site. The
Hamiltonian we solve is $H= H_{K}+ H_{I}$
where \begin{equation}\begin{split} H_{K}&=-\sum_{\langle ij\rangle
      m\sigma}t_{m}c^+_{jm\sigma}c_{im\sigma}-
    t_\perp\sum_{i\sigma}c^+_{i1\sigma}c_{i2\sigma}\\
    H_{I} &= U \sum_{mi} n_{mi\uparrow} n_{mi\downarrow} +
    \sum_{\sigma}\big[(U-2J_z) n_{1i\sigma} n_{2i\bar{\sigma}} +\\&+
    (U-3J_z)n_{1i\sigma}n_{2i\sigma}\big].
\label{eq:hami} \end{split}\end{equation}

Bands which are away from the Fermi level are not expected to be
significantly renormalized by correlations, and we save on
computational effort by ignoring them in the DMFT calculation. As an
impurity solver for the DMFT equations we employed the weak-coupling
continuous time Quantum Monte Carlo. Our results are for $T=0.025$~eV,
$U=1.6$~eV and $J_z=U/8$. This choice of interaction parameters was
guided by recent quantum chemistry~\cite{Giovanetti2011} and
constrained random phase approximation (cRPA)
estimates~\cite{Nomura2012}. Note that for K$_3$picene, we find the
metal-insulator transition between $U=0.6$~eV (metallic) and
$U=0.8$~eV (insulating). The analytic continuation of the converged
impurity Green's function to real frequencies was performed with the
maximum entropy method.  The resulting spectral function for these two
orbitals is shown in Fig.~\ref{theory}, embedded into the DFT-DOS of
the non-interacting bands. 

\begin{figure}[b] \includegraphics[width=0.45\textwidth]{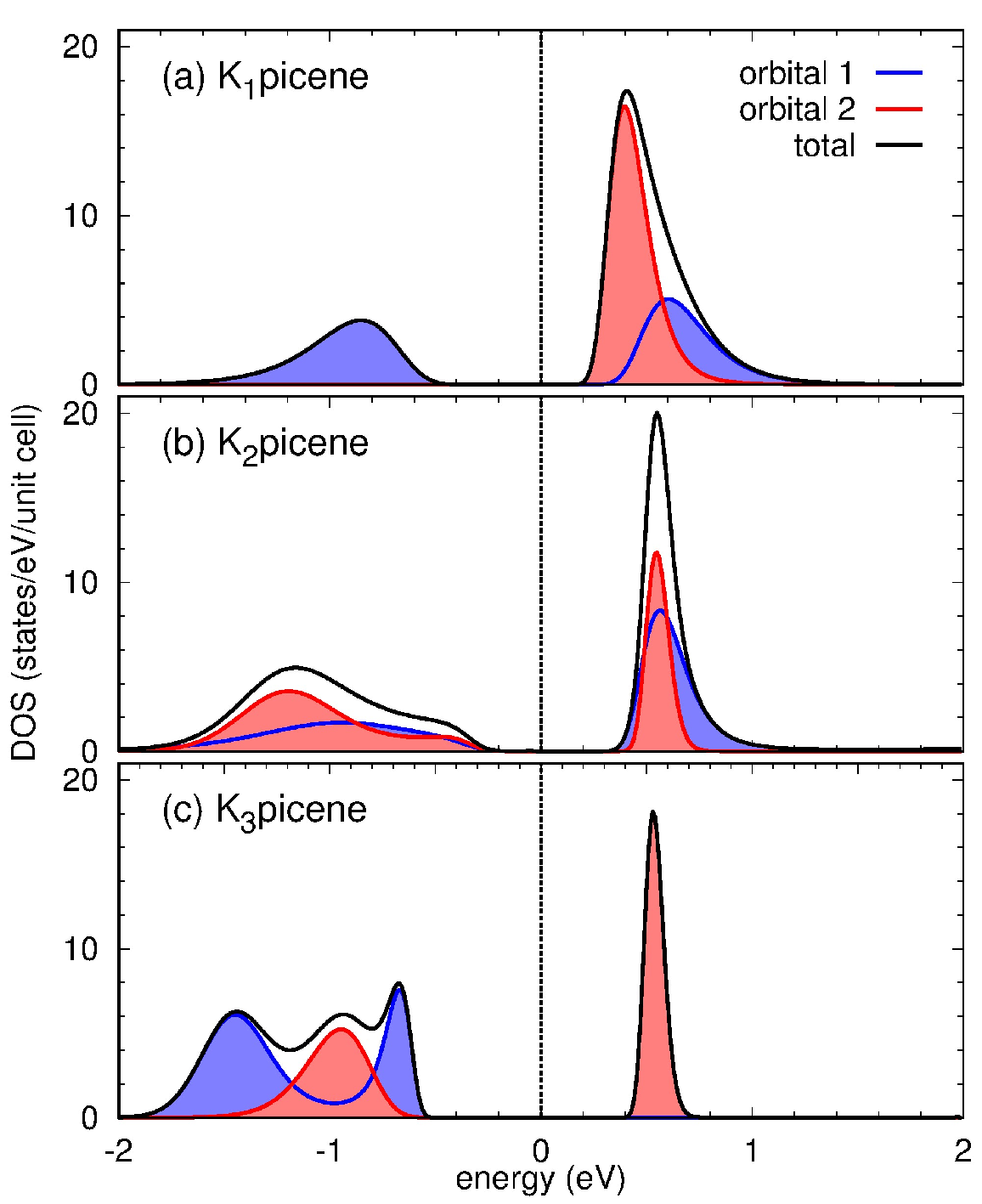}
\caption{Orbital resolved and total DMFT spectral functions of (a) K$_1$picene,
(b) K$_2$picene and (c) K$_3$picene.}\label{theory1} \end{figure}

Clearly, the DMFT spectral functions show a Mott gap not only for
doping levels of $x=1$ and $x=3$, but also for $x=2$ which in bare DFT
is a band insulator with a small gap. A closer look at the DOS
(Fig.~\ref{theory1}~(b)) of K$_2$picene shows that -- contrary to the
expectation of a filled picene LUMO derived band and an empty picene
LUMO+1 derived band -- a Mott gap opens in \textit{both}
orbitals. Interestingly, energy positions and gap size do not seem to
depend much on the actual doping level, except for the spectral weight
transfer from upper to lower Hubbard bands, and correspond favorably
with the experimental spectra, except for some fine structure at
higher $x$ not resolved in photoemission. The origin of the Mott gap for 
K$_x$picene with $x=1,2,3$ lies not only in the on-site Coulomb interaction
 $U$ and the Hund's coupling $J_z$ but also in the interorbital hybridization $t_{\perp}$. We would like to note that
 our calculations did neither include  electron-phonon effects, which may reduce the Hubbard interaction, as was shown by 
Sangiovanni {\it et al.}~\cite{Sangiovanni2005}, nor magnetic ordering, where the possibility of
a Mott insulator in the antiferromagnetic state could be investigated, as was found by the authors
of Refs.~\onlinecite{Giovanetti2011,Kim2011} by performing {\it ab initio} DFT
calculations. This will be a subject of future work.

Both experiment and theory thus arrive at the same conclusion that K$_x$picene
films are (Mott-)insulating for all compositions with integer $x$, including the nominally 
superconducting phase $x=3$. This leaves two open questions: (1) What is the 
electronic groundstate for \textit{non-integer} $x$, and (2) how can the insulating 
behavior of the $x=3$ phase be reconciled with the reports on superconductivity?

Concerning the first question we note that in a Mott-insulator deviation from
integer band-filling is normally expected to lead to a metallic phase with
electron- or hole-type carriers \cite{Imada1998}. Note, however, that the doping
concentration determined in our photoemission experiments is a macroscopic quantity, averaged
over the extension of the photon spot ($\sim1$ mm), and may be composed of
phase-separated integer-filled domains. In fact, a recent Raman study
\cite{Kambe2012} has observed that K$_x$picene decomposes into $x=0$, 2 and 3
phases, while a defined $x=1$ signal could not be detected, indicating that this
phase is not stable. One can thus understand the photoemission spectra as a
superposition of different integer $x$ phases. Because the shape and gapped
behavior of the corresponding theoretical spectra is very similar, the main
doping effect will consist in the appearance and increasing spectral weight of
the LUMO and LUMO+1 derived lower Hubbard band, as indeed reflected by the
behavior of feature F$_1$ in our experimental spectra.

The second question is more puzzling. Our present results are in line with the 
general observation that alkali metal-doped organic semiconductors are mostly 
insulating \cite{Koch2003, Fuentes2006}, with K-doped fullerene (K$_3$C$
_{60}$) being one of the rare exceptions \cite{Poirier1994}. On the other hand, in the original 
paper by Mitsuhashi {\it et al.} it has been reported that in K$_{2.9}$picene the 
diamagnetic (i.e., superconducting) phase is reached from a Pauli paramagnetic, hence \textit{metallic} phase
\cite{Mitsuhashi2010}. This is also supported by the resistivity measurements by Teranishi {\it et al.}~\cite{Teranishi2013}.
Reconciling our room-temperature results with these
observations would require an insulator-to-metal transition somewhere between 
300 and 25 K (the highest measured temperature in Ref.~\onlinecite{Mitsuhashi2010}).
The Mott-Hubbard scenario derived from our DFT+DMFT calculations does not provide 
any mechanism for such a transition but would rather stabilize the gap towards low temperatures. 

Another important aspect is that so far all superconducting signatures have been obtained from 
intercalated compact picene powder \cite{Mitsuhashi2010, Kambe2012}, 
in contrast to the \textit{in vacuo} grown thin films used here. It is possible that both sets of 
samples differ in subtle structural details accounting for the conflicting findings. On the other hand, 
the DFT+DMFT calculations are based on a DFT-relaxed \textit{bulk} structure which one would 
expect to be relevant for the polycrystalline powder, at least for the interior of the powder grains. 
It is thus tempting to speculate about the possibility of the superconducting phase actually being 
located at the grain boundaries, which may have structural and/or electronic properties different from the bulk.      
Certainly more clarifying work needs to be done in this direction.

In summary, our combined photoemission and DFT+DMFT study finds that K$_x$picene
is a correlated Mott insulator for any integer doping concentration $x=1$, 2 and 3. The insulating 
behavior of samples with macroscopic non-integer filling observed in the experiment is attributed to 
microscopic phase separation in line with recent Raman results. The apparent spectral weight at or even 
above the Fermi level experimentally observed for low doping ($x\lesssim1$) is attributed to a 
lack of thermodynamical equilibrium, resulting from weak intermolecular van der Waals interactions. 
The resulting discrepancy between the Mott-insulating behavior of particularly K$_3$picene found 
here and the recent reports of  superconducting signatures~\cite{Mitsuhashi2010,Teranishi2013} calls for further 
investigations.

\begin{acknowledgments} We wish to thank N.H. Hansen and J. Pflaum (both at University 
of W\"urzburg) for their helpful technical assistance and Kristjan Haule for very useful discussions. This work was supported
by the Deutsche Forschungsgemeinschaft (FOR1346) and the Helmholtz Association (HA216/EMMI). \end{acknowledgments}

\newpage
~~~
\newpage

\renewcommand{\thefigure}{S\arabic{figure}}
\setcounter{figure}{0}
\renewcommand{\thetable}{S\Roman{table}}
\setcounter{table}{0}

\section{Supplemental Material}

\subsection{Structural characterization}

\begin{figure*}[h]
 \includegraphics[width=0.95\textwidth]{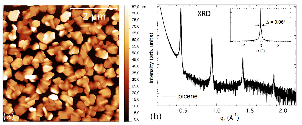}
 \caption{(a) 5$\times$5\,$\mu$m$^2$ AFM image of a nominally 30 nm (QCM) thick picene film showing azimuthally disordered islands. (b) X-ray diffraction pattern and rocking curve (inset) of a pristine picene film. The film is highly ordered along the c-axis and shows only a small mosaicity.}\label{AFM}
 \end{figure*}

The typical average thickness of the \textit{in vacuo} grown picene
films as monitored by a quartz crystal microbalance (QCM) was 30~nm.
The actual film topography as measured by atomic force microscopy
(AFM) shows that the film consists of picene islands with
typical lateral dimensions of several 100 nm and heights of up to 80
nm (Fig.~\ref{AFM}~(a)). Such 3D island growth mode is expected on an
inert substrate (here: chemically cleaned and oxidized Si(100)).
While the individual islands are crystalline, the overall film appears
azimuthally disordered with well-developed texture along the c-axis,
i.e., perpendicular to the surface, as inferred from the sharp
Bragg peaks in the x-ray diffractogram (XRD) in Fig.~\ref{AFM}~(b). The resulting c-axis parameter of
13.51\AA \,confirms that the molecules grow in a standing mode.

\subsection{Doping concentration}

The atomic K concentration $x$ of the doped picene films (K$_x$picene)
has been determined from the intensity ratio of the C~1\textit{s} and
K~2\textit{p} photoemission core level spectra, taking into account
the relative photoionization cross sections.  The latter is
essentially an atomic property and has experimentally been calibrated
by independent photoemission measurements of K-intercalated fullerenes
(K$_3$C$_{60}$), which is an easily reproducible compound of
well-defined stoichiometry.  Note that the relative cross-section
ratio $\sigma_{C 1s}/\sigma_{K 2p}$ thus determined is about 30~\%
smaller than the value obtained from the theoretical cross-sections of
Yeh and Lindau \cite{SYeh1985}. This implies that previous
photoemission studies which use the calculated values overestimate the
K concentration by the same amount.

\subsection{Non-equilibrium effects}

\begin{figure*}[t]
 \includegraphics[width=0.95\textwidth]{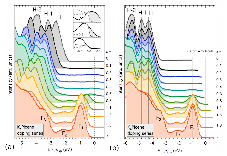}
 \caption{Photoemission spectra of K$_x$picene in a range of
   $\sim$\,5\,eV of the HOMO region for various K-doping
   concentrations. Spectra are shown for $x =0 \dots 3$.  (a) energies
   as measured, (b) energetically aligned on the spectral features of
   the $x=3$ compound.}\label{series}
\end{figure*}

Figure~\ref{series}(a) shows a series of valence band spectra as
measured for a full range of K concentrations from $x=0$ to 3. The
main effect at low doping ($x\lesssim1$) is a shift of the overall
spectrum away from the experimental Fermi energy $\mu_\text{exp}$ (which
corresponds to the kinetic photoelectron energy where a metal in
electrical contact with the sample would display its Fermi edge) and
the appearance of a small peak (F$_1$) in the former band gap. Closer
inspection reveals that most of its spectral weight is actually
located well \textit{above} the Fermi energy for $x=0.1$ and only
gradually moves below $\mu_\text{exp}$ with larger doping, until the peak
finally converges to a constant binding energy for $x\gtrsim1$,
resulting in a gapped spectrum. This anomalous behavior clearly signals
a non-equilibrium situation where a homogeneous chemical potential
cannot be defined in the film-substrate system in the early doping
stages. A possible explanation by a surface photovoltage (SPV), often
encountered in metal-semiconductor contacts, can be ruled out, because
we do not observe the characteristic temperature and intensity
dependence of the SPV \cite{SAlonso1990,SHecht1990,SHorn1992,SMoreno2000}.

A detailed analysis of identical behavior in another alkali metal-doped
organic semiconductor (\textit{para}-sexiphenyl) has demonstrated that
it can be traced back to the inability of a purely van der
Waals-coupled molecular system to establish an equilibrium chemical
potential throughout the film \cite{SKoch2003}. In such a situation the
energy alignment refers rather to the vacuum level and thus
depends on work functions. Charge fluctuations induced by alkali-metal
doping will restore thermodynamical equilibrium. Note however, that
our potassium deposition method leads to a concentration gradient
which decays from the surface into the film. Low K dosage then leads
to a doped K$_x$picene surface layer with little or even no K present
deeper in the film, maintaining the non-equilibrium situation
described above. For higher concentrations the film will become doped
throughout, hence restoring a well-defined chemical potential. The
observed effect should be less pronounced for thinner films, as is
indeed the case for the photoemission results of Mahns {\it et al.}
\cite{SMahns2012}.

We thus take our $x=3$ data as an intrinsic equilibrium spectrum and
have energetically realigned all other spectra at its spectral
features in Fig.~\ref{series}(b), showing gapped behavior for all
concentrations..  The observation of spectral weight near the Fermi
level reported by Okazaki {\it et al.} \cite{SOkazaki2010} for a
nominal doping of $x=1$ is attributed to the non-equilibrium behavior
of peak F$_1$, which also explains why a clear Fermi edge could not be
identified in that study.

\section{Structure prediction}

\begin{figure*}[t]
 \includegraphics[width=0.95\textwidth]{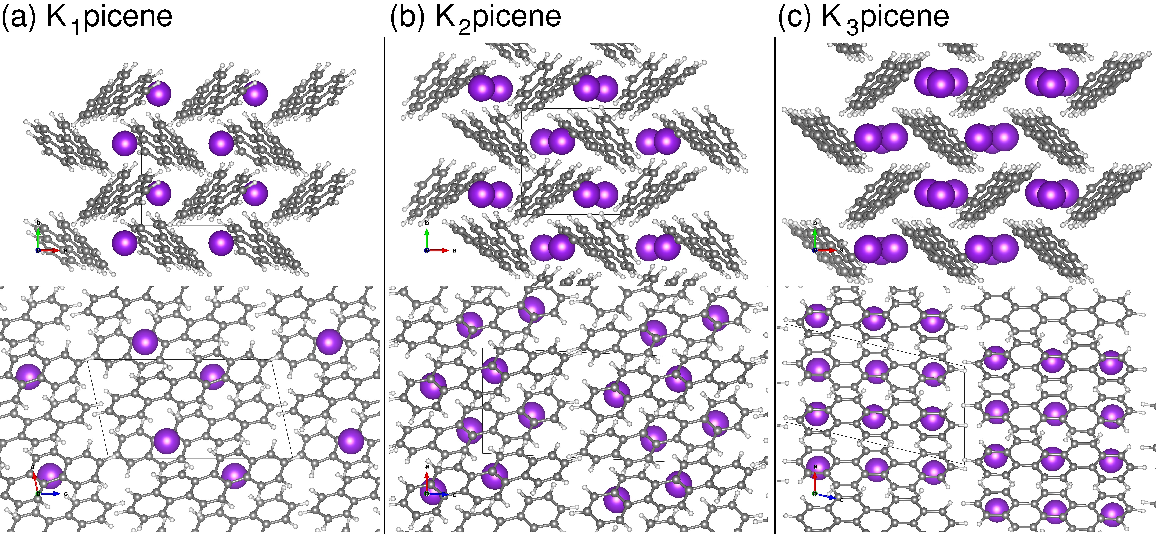}
 \caption{Predicted structures of K$_x$picene with $x=1,2,3$.}\label{fig:struct}
\end{figure*}

The relaxation of K$_x$picene candidate structures was performed using
Fast Inertial Relaxation Engine (FIRE) algorithm \cite{SBitzek2006},
employing Vienna Ab-initio Simulations Package (VASP)
\cite{SKresse1993,SKresse1996} with the Projector Augmented Wave basis
(PAW) \cite{SBloechl1994,SKresse1999} in the Generalized Gradient
Approximation (GGA) as a force and stress tensor calculator with
plane-wave cutoff of 400~eV and Brillouin zone sampled by
Monkhorst-Pack $k$-point mesh of $5\times 5\times 5$
$k$-points. Structures were relaxed until the maximal force acting on
ions and maximal stress tensor component fell bellow 0.01~eV/A and
0.01~GPa, respectively. Fig.~\ref{fig:struct} shows the structures we
obtained. We have compared our structures to theoretically obtained
lattice parameters published in Ref.~\onlinecite{SKosugi2011}. Obtained
equilibrium volumes agree with our results within 2\% in all cases,
but there are some discrepancies in lattice parameters and
angles. Without detailed knowledge of the structures in
Ref.~\onlinecite{SKosugi2011} it is not possible to say anything
further on the degree of agreement.

\section{Details of dynamical mean field theory calculations}

We align Fermi levels of DFT spectra with
active bands removed and DMFT spectral functions at $E=0$. While the
DFT Fermi level is known for K$_1$picene and K$_3$picene, as they are
metallic, and for K$_2$picene with an error bar given by the tiny
energy gap of 60 meV, the DMFT chemical potential follows from the
demand that the filling is unchanged compared to the DFT calculation.

\end{document}